\documentclass{JHEP3}

\usepackage{epsfig}

\makeatletter\@addtoreset{equation}{section}\makeatother

\makeatletter

\@addtoreset{equation}{section}
\makeatother

\newcommand{\be}{\begin{equation}}

\newcommand{\ee}{\end{equation}}

\newcommand{\bea}{\begin{eqnarray}}

\newcommand{\eea}{\end{eqnarray}}

\newcommand{\xx}{\nonumber\\}

\newcommand{\ct}{\cite}

\newcommand{\la}{\label}

\newcommand{\eq}[1]{(\ref{#1})}

\def\CL{{\cal L}}

\def\CO{{\cal O}}

\def\IR{{\hbox{{\rm I}\kern-.2em\hbox{\rm R}}}}
\def\IB{{\hbox{{\rm I}\kern-.2em\hbox{\rm B}}}}
\def\IN{{\hbox{{\rm I}\kern-.2em\hbox{\rm N}}}}
\def\IC{\,\,{\hbox{{\rm I}\kern-.59em\hbox{\bf C}}}}
\def\IZ{{\hbox{{\rm Z}\kern-.4em\hbox{\rm Z}}}}
\def\IP{{\hbox{{\rm I}\kern-.2em\hbox{\rm P}}}}
\def\IH{{\hbox{{\rm I}\kern-.4em\hbox{\rm H}}}}
\def\ID{{\hbox{{\rm I}\kern-.2em\hbox{\rm D}}}}

\def\half{\frac{1}{2}}

\def\p{\partial}

\def\det{{\rm det}}

\preprint{SNUTP 03-027 \\
{\tt hep-th/0312103}}

\author{Rabin Banerjee$^{\, a \,}$\footnote{rabin@newton.skku.ac.kr} \footnote{On leave from
S. N. Bose National Center for Basic Sciences, Calcutta, India;
rabin@bose.res.in},
Choonkyu Lee$^{\, b \,}$\footnote{cklee@phya.snu.ac.kr}
and Hyun Seok Yang$^{\, b \,}$\footnote{hsyang@phya.snu.ac.kr}
\\
${}^a${\sl BK21 Physics Research Division and Institute of Basic Science, \\
Sungkyunkwan University, Suwon 440-746, Republic of Korea} \\
${}^b${\sl School of Physics and Center for Theoretical Physics, \\
Seoul National University, Seoul 151-747, Republic of Korea}}

\title{Seiberg-Witten-type Maps for Currents and Energy-Momentum Tensors
in Noncommutative Gauge Theories}

\abstract{We derive maps relating the currents and energy-momentum tensors
in noncommutative (NC) gauge theories with their commutative
equivalents. Some uses of these maps are discussed. Especially, in
NC electrodynamics, we obtain a generalization of the Lorentz
force law. Also, the same map for anomalous currents relates the
Adler-Bell-Jackiw type NC covariant anomaly with the standard
commutative-theory anomaly. For the particular case of two
dimensions, we discuss the implications of these maps for the
Sugawara-type energy-momentum tensor.}


\begin{document}

\renewcommand{\thefootnote}{\arabic{footnote}}
\setcounter{footnote}{0}

\section{Introduction}

The Seiberg-Witten (SW) map, first formulated in \ct{sw},
establishes an equivalence between noncommutative (NC) gauge
theories and conventional gauge theories defined on ordinary
(commutative) space. Consequently it becomes feasible to discuss
several features of NC gauge theories in their commutative
equivalents, thereby making the former more tractable. So far this
analysis has been confined to source-free theories since the
original map was given for the gauge potentials. In order to
discuss NC gauge theories with sources it is therefore essential
to have a corresponding map for the sources, which is otherwise
lacking. One of the objectives of this paper is to provide such
maps and also for the energy-momentum (EM) tensors.

A vexing issue is the apparent lack of agreement in the results
obtained by first applying the map on the NC action with the
source term added and then analyzing the equations of motion or,
alternatively, by first obtaining the equations of motion in the
NC version and then exploiting the map. These points were raised
and discussed (for the source-free case) in \ct{kruglov,grimstrup,das} in various contexts.
During the course of our analysis we show that, with proper
interpretation, all disagreement or ambiguities are ironed out.

As stated earlier, we derive a map for the sources or the
currents. This is a general result which can be expressed in a
closed form. The map is explicitly worked out for the first
nontrivial
order in $\theta$, which is the NC parameter. It is then used to
relate the usual gauge invariant Adler-Bell-Jackiw (ABJ) anomaly
\ct{abj} in the commutative case with the star gauge covariant
anomaly in the NC theory. The interplay of anomaly with gauge
invariance (or covariance) is also discussed. We then extend the
analysis to provide a map for the energy-momentum tensors in the
two descriptions, clarifying en route some subtleties in their
definition. Along with the equations of motion in NC
electrodynamics with sources, this yields the Lorentz force law.

It is our belief that the SW-type maps for the currents and EM
tensors, aside from such familiar maps for the gauge and matter
fields, deserve attention on its own right. Particularly, it
allows one to discuss various physical aspects, irrespectively of
either the detailed form or the SW maps for the matter fields. In
this connection, we also discuss briefly their implications on an
intriguing Sugawara-type formulation where the EM tensor is
expressed sorely in terms of the currents. Compared to the
dimension independent analysis in the rest of the paper, this part
is confined to two dimensions.

In section 2, we derive the map for currents and anomalies.
Section 3 contains the corresponding analysis for EM tensor and
the derivation of the Lorentz force law in NC electrodynamics.
Section 4 has, apart from the concluding remarks, a Sugawara-type
construction in two dimensions which is also compatible with the
results obtained in the previous sections.

\section{Map for Currents and Anomalies}

Here we derive a mapping of the currents in the NC and commutative
descriptions. Also, this will be used to provide a map between the
different anomalies. First, an algebraic approach is discussed
where the results are given to the first order in the NC
parameter, $\theta$. This will be subsequently generalized, in a
dynamical approach, to all orders in $\theta$.

\subsection{Algebraic approach}

The original map \ct{sw,wess1,wess2} relating the gauge potentials and field
tensors in NC $U(1)$ gauge theory \footnote{Variables in the NC space
are distinguished from their conventional counterparts by a
caret.},
\begin{eqnarray}\label{sw-a}
&& \widehat{A}_\mu = A_\mu - \frac{1}{2}\theta^{\alpha\beta} A_\alpha
(\partial_\beta A_\mu + F_{\beta\mu}) + \CO(\theta^2), \\
\label{sw-f}
&& \widehat{F}_{\mu \nu} = F_{\mu\nu} + \theta^{\alpha\beta}
(F_{\mu\alpha} F_{\nu \beta} - A_\alpha \partial_\beta
F_{\mu\nu}) + \CO(\theta^2)
\end{eqnarray}
was obtained algebraically so that the stability of gauge
transformations,
\bea \la{swmap-nca}
&& \widehat{\delta}_{\widehat{\lambda}} \widehat{A}_\mu \equiv
\widehat{D}_\mu \star \widehat{\lambda} = \p_\mu \widehat{\lambda} - i
\widehat{A}_\mu \star \widehat{\lambda} + i \widehat{\lambda}
\star \widehat{A}_\mu = \p_\mu \widehat{\lambda} +
\theta^{\alpha\beta} \p_\alpha \widehat{A}_\mu \p_\beta
\widehat{\lambda} + \CO(\theta^2), \\
\la{swmap-ca}
&& \delta_\lambda A_\mu \equiv \p_\mu \lambda,
\eea
may be insured by a further map among the gauge parameters
\begin{equation}\label{swp-para}
    \widehat{\lambda} = \lambda + \frac{1}{2} \theta^{\alpha\beta}
    \p_\alpha \lambda A_\beta +  \CO(\theta^2).
\end{equation}
It may be noted that \eq{sw-f} is a consequence of
\eq{sw-a}, following from the basic definitions
\bea \la{nc-f}
\widehat{F}_{\mu\nu} &\equiv& \p_\mu \widehat{A}_\nu -  \p_\nu \widehat{A}_\mu
- i \widehat{A}_\mu \star \widehat{A}_\nu + i \widehat{A}_\nu
\star \widehat{A}_\mu \xx
&=& \p_\mu \widehat{A}_\nu -  \p_\nu \widehat{A}_\mu
+ \theta^{\alpha\beta}\p_\alpha \widehat{A}_\mu \p_\beta \widehat{A}_\nu
+  \CO(\theta^2)
\eea
and
\begin{equation}\label{c-f}
F_{\mu\nu} \equiv \p_\mu A_\nu -  \p_\nu A_\mu,
\end{equation}
so that, whereas $F_{\mu\nu}$ is gauge invariant,
$\widehat{F}_{\mu\nu}$ transforms covariantly under the star gauge
transformation,
\begin{equation}\label{star-gtf}
\widehat{\delta}_{\widehat{\lambda}} \widehat{F}_{\mu\nu} \equiv - i
\widehat{F}_{\mu\nu} \star \widehat{\lambda} + i \widehat{\lambda}
\star \widehat{F}_{\mu\nu} = \theta^{\alpha\beta} \p_\alpha \widehat{F}_{\mu\nu}
\p_\beta \widehat{\lambda} + \CO(\theta^2).
\end{equation}

The proposed map among the currents $\widehat{J}^\mu$ and $J^\mu$
is now obtained under the following two conditions: the current
$J^\mu$ is gauge invariant and satisfies the ordinary conservation
law $\p_\mu J^\mu=0$, while the current $\widehat{J}^\mu$
transforms covariantly and satisfies the covariant conservation
law $\widehat{D}_\mu \star \widehat{J}^\mu=0$. Up to
$\CO(\theta)$, the stability under gauge transformations is easily
attained by mimicking the map \eq{sw-f} among the field tensors,
\begin{equation}\label{swder-j}
    \widehat{J}^\mu = J^\mu - \theta^{\alpha\beta} A_\alpha
    \p_\beta J^\mu + \cdots,
\end{equation}
where the ellipses indicate the freedom of adding more terms that
are invariant under ordinary gauge transformations. It is clear
that the most general structure is given by
\begin{equation}\label{gener-ans-j}
\widehat{J}^\mu = J^\mu - \theta^{\alpha\beta} A_\alpha
    \p_\beta J^\mu + c_1 \theta^{\mu \alpha} F_{\alpha\beta}
  J^\beta + c_2 \theta^{\alpha\beta} F_{\alpha\beta} J^\mu
  + c_3 \theta^{\alpha\beta} {F_\alpha}^\mu J_\beta,
\end{equation}
where $c_1, c_2$, and $c_3$ are undetermined coefficients. Next,
demanding the simultaneous conservation $\widehat{D}_\mu \star \widehat{J}^\mu
=\p_\mu J^\mu=0$ immediately fixes $c_1 = 2c_2 = 1$ and $c_3 = 0$
so that,
\bea \label{current}
  \widehat{J}^\mu  &=& J^\mu - \theta^{\alpha\beta} A_\alpha
  \partial_\beta J^\mu + \theta^{\mu \alpha} F_{\alpha\beta}
  J^\beta +\frac{1}{2} \theta^{\alpha\beta} F_{\alpha\beta} J^\mu \xx
  &=& J^\mu + (\theta F J)^\mu + \p_\alpha (\theta^{\alpha\beta} A_\beta J^\mu)
\eea
where an obvious matrix notation has been introduced.

This yields the cherished map among the currents valid up to
$\CO(\theta)$. Observe that the derivation is based on general
gauge transformation properties. The explicit structure of neither
$\widehat{J}^\mu$ nor $J^\mu$ need to be specified. If any one of
these is known, the other is determined through the map \eq{current}
or its inverse
\begin{equation}\label{isw-j}
J^\mu  = \widehat{J}^\mu - (\theta \widehat{F} \widehat{J})^\mu
- \p_\alpha (\theta^{\alpha\beta} \widehat{A}_\beta \widehat{J}^\mu).
\end{equation}
We now present a dynamical treatment which generalizes the above
results, apart from precisely specifying the currents.

\subsection{Dynamical approach}

Let the noncommutative action be defined as
\begin{eqnarray} \label{lee-ncqed}
    \widehat{S}(\widehat{A}, \widehat{\psi}) &=&  -\frac{1}{4}
    \int d^4 x \widehat{F}_{\mu\nu} \star
    \widehat{F}^{\mu\nu} + \widehat{S}_M (\widehat{\psi}, \widehat{A}) \nonumber \\
& \equiv & \widehat{S}_{\rm{ph}} (\widehat{A})
+ \widehat{S}_M (\widehat{\psi}, \widehat{A})
\end{eqnarray}
where the pure gauge term has been isolated in the ``photonic"
piece $ \widehat{S}_{\rm{ph}} (\widehat{A})$. The
charged matter fields are denoted by $\widehat{\psi}_\alpha$.
Then the equation of motion for $\widehat{A}_\mu$ is
\begin{equation}\label{lee-eom}
    \frac{\delta \widehat{S}_{\rm{ph}} (\widehat{A})}
    {\delta \widehat{A}_\mu}
= \widehat{D}_\nu \star \widehat{F}^{\nu \mu} = - \widehat{J}^\mu
\end{equation}
where
\begin{equation}\label{lee-def-j}
    \widehat{J}^\mu := \frac{\delta \widehat{S}_M (\widehat{\psi},\widehat{A})}
    {\delta \widehat{A}_\mu}|_{\widehat{\psi}}.
\end{equation}
Here, thanks to the equation of motion satisfied by
$\widehat{\psi}_ \alpha$, $\widehat{J}^\mu$ will satisfy the
covariant conservation law
\begin{equation}\label{lee-claw}
\widehat{D}_\mu \star \widehat{J}^\mu = 0.
\end{equation}
This may also be seen by taking the covariant derivative on both
sides of Eq. \eq{lee-eom}. The same equation also shows that
$\widehat{J}^\mu$ transforms covariantly under the star gauge
transformations. Clearly therefore, this $\widehat{J}^\mu$ is
similar to the one considered in section 2.1.

Now we rewrite the action (\ref{lee-ncqed}) using the SW map to
obtain a $U(1)$ gauge invariant action defined on commutative
space \ct{bichl,carroll} \footnote{A SW map for the matter sector
is also necessary for this transition but its explicit structure
is inconsequential for this analysis.},
\begin{equation}\label{lee-sw-action}
 \widehat{S}(\widehat{A}, \widehat{\psi})|_{\rm{SW \; map}}
 := S_{\rm{ph}}(A) + S_M (\psi, A)
\end{equation}
where $S_{\rm{ph}}(A)$ contains all terms involving $A^\mu$ only
and is given by
\begin{equation}\label{lee-dqed}
S_{\rm{ph}}(A) = \int d^4 x \Bigl[ -\frac{1}{4}F_{\mu\nu}F^{\mu\nu}
 - \frac{1}{2}\theta^{\alpha\beta}(F_{\mu\alpha}F_{\nu\beta}F^{\mu\nu}
 - \frac{1}{4} F_{\alpha\beta}F_{\mu\nu}F^{\mu\nu}) + \CO(\theta^2)
 \Bigr].
\end{equation}
In fact this action, modulo constant terms, is the expansion of
the Born-Infeld action up to order $\CO(F^3)$ (with $2\pi \alpha^\prime =1$) \cite{sw},
\begin{equation}\label{bi}
    S_{BI} = \int d^4x \sqrt{-\det(\eta_{\mu\nu} - \theta_{\mu\nu} +
    F_{\mu\nu})}.
\end{equation}
Now from Eq. \eq{lee-dqed} the gauge invariant equation of motion is
obtained,
\begin{equation}\label{lee-gi-eom}
 \frac{\delta S_{\rm{ph}} (A)}
    {\delta A_\mu (x)} = - J^\mu (x)
\end{equation}
where
\begin{equation}\label{lee-def-invj}
J^\mu (x) := \frac{\delta S_M (\psi,A)}
    {\delta A_\mu (x)}|_{\psi}
\end{equation}
which in general contains $\theta$-dependent terms.
Again, thanks to the equation of motion satisfied by
$\psi_ \alpha$, $J^\mu$ now satisfies the ordinary
conservation law,
\begin{equation}\label{lee-iclaw}
    \partial_\mu J^\mu = 0.
\end{equation}
The same result is also inferred from the gauge invariance of
Eqs. \eq{lee-dqed} and \eq{lee-gi-eom}. This current, therefore, is
similar to $J^\mu$ introduced in section 2.1.

It is now possible to obtain a relation between $\widehat{J}^\mu$ and
$J^\mu$ by noticing that
\begin{eqnarray}\label{lee-gsw}
 \widehat{J}^\mu (x) &=& \frac{\delta \widehat{S}_M (\widehat{\psi},\widehat{A})}
    {\delta \widehat{A}_\mu}|_{\widehat{\psi}} \nonumber \\
    &\stackrel{\rm{SW \; map}}{=}& \int d^4 y \Bigl[ \frac{\delta S_M (A, \psi)}
    {\delta A_\nu (y)}|_{\psi} \frac{\delta A_\nu (y)}{\delta \widehat{A}_\mu (x)}
    + \frac{\delta S_M (A, \psi)}
    {\delta \psi_\alpha (y)}|_{A} \frac{\delta \psi_\alpha (y)}
    {\delta \widehat{A}_\mu (x)} \Bigr] \xx
    &=& \int d^4 y J^\nu (y) \frac{\delta A_\nu (y)}
    {\delta \widehat{A}_\mu (x)},
\end{eqnarray}
where the second term in the second line was dropped on using
the equation of motion for $\psi_\alpha $.
Eq. \eq{lee-gsw} yields the general form of the map between the
currents. Although it is displayed for four dimensions, the result
is obviously valid for any dimensions.

As a simple yet nontrivial check, we now reproduce the
$\CO(\theta)$ result \eq{current}, starting from Eq. \eq{lee-gsw}.
From Eq. \eq{sw-a} it follows that
\be \la{func-der}
\frac{\delta}{\delta \widehat{A}_\mu (x)} =
\frac{\delta}{\delta A_\mu (x)} + \int d^4 y \theta^{\alpha\beta}
\frac{\delta}{\delta A_\mu (x)} \Bigl( A_\alpha(y) \p_\beta
A_\lambda(y) - \half  A_\alpha(y) \p_\lambda
A_\beta(y) \Bigr) \frac{\delta}{\delta A_\lambda (y)}+
\CO(\theta^2).
\ee
Using this to evaluate the functional derivative in Eq. \eq{lee-gsw}
immediately leads to Eq. \eq{current} where, at an intermediate step,
the current conservation \eq{lee-iclaw} has been used.

The map (\ref{lee-gsw}) is also consistent with the observation,
\begin{eqnarray}\label{lee-sw-eom}
 - \widehat{J}^\mu (x) &=&  \widehat{D}_\nu \star \widehat{F}^{\nu
 \mu} = \frac{\delta \widehat{S}_{\rm{ph}} (\widehat{A})}
    {\delta \widehat{A}_\mu} \nonumber \\
    &\stackrel{\rm{SW \; map}}{=}& \int d^4 y \frac{\delta S_{\rm{ph}} (A)}
    {\delta A_\nu (y)} \frac{\delta A_\nu (y)}{\delta \widehat{A}_\mu (x)}
    =  - \int d^4 y J^\nu (y) \frac{\delta A_\nu (y)}
    {\delta \widehat{A}_\mu (x)}.
\end{eqnarray}
It is also clear that the effective (non-linear) Maxwell
equation with (gauge-invariant) source $J^\mu$ is
naturally identified with the expression (\ref{lee-gi-eom}). Note,
however, that this is in general different from the stationary
condition obtained by applying the SW map to an action
\begin{equation}\label{nc-j}
    \widehat{S}_{J} = \int d^4 x \Bigl[-\frac{1}{4}\widehat{F}_{\mu\nu} \star
    \widehat{F}^{\mu\nu} + \widehat{A}_\mu \star \widehat{J}^\mu
    \Bigr],
\end{equation}
although it also leads to the equation of motion \eq{lee-eom}.
The discrepancy  arises because the source term in Eq. \eq{nc-j} is not gauge
invariant under NC $U(1)$ gauge transformations (with $\widehat{J}^\mu$
in the adjoint representation) so that the application of the SW
map becomes meaningless. It is only after the inclusion of the
full matter sector that gauge invariance is restored, leading to
our original action \eq{lee-ncqed}.

As another consistency check on the construction \eq{lee-gsw} or
\eq{lee-sw-eom}, observe that the latter leads to an identity if
everything is expressed in terms of the gauge potentials,
\begin{equation}\label{add-sw-eom}
 \widehat{D}_\nu \star \widehat{F}^{\nu \mu} \stackrel{\rm{SW \; map}}{=}
 \int d^4 y \frac{\delta S_{\rm{ph}} (A)}
 {\delta A_\nu (y)} \frac{\delta A_\nu (y)}{\delta \widehat{A}_\mu (x)},
\end{equation}
where
\begin{eqnarray}\label{deom}
\frac{\delta S_{\rm{ph}} (A)}{\delta A_\nu } &=& \partial_\mu F^{\mu\nu}
- \frac{1}{2} \theta^{\alpha\beta} \partial_\mu
    (F_{\alpha\beta} F^{\mu\nu}) -  \frac{1}{4}\theta^{\mu\nu} \partial_\mu
    (F_{\alpha\beta} F^{\alpha\beta}) + \theta^{\nu \alpha} \partial_\mu
    (F_{\alpha\beta} F^{\beta \mu}) \nonumber \\
&& - \theta^{\mu \alpha} \partial_\mu (F_{\alpha\beta} F^{\beta \nu})
+ \theta_{\alpha \beta} \partial_\mu
    (F^{\alpha\mu} F^{\beta\nu}) + \CO(\theta^2)
\end{eqnarray}
is obtained from Eq. \eq{lee-dqed}. Up to $\CO(\theta)$ the left-hand
side of Eq. \eq{add-sw-eom} can be computed from a direct
application of the SW map \eq{sw-a}-\eq{sw-f}, leading to,
\begin{equation}\label{eom-gauge}
\widehat{D}_\mu \star \widehat{F}^{\mu\nu} \stackrel{\rm{SW \; map}}{=}
\partial_\mu F^{\mu \nu}
- \theta^{\alpha\beta} A_\alpha \partial_\beta \partial_\mu F^{\mu \nu}
+ \theta_{\alpha\beta} \partial_\mu (F^{\alpha \mu} F^{\beta \nu})
+ \theta^{\alpha\mu} F_{\alpha\beta}
  \partial_\mu F^{\beta\nu}.
\end{equation}
The right-hand side of Eq. \eq{add-sw-eom} is next computed using
Eqs. \eq{deom} and \eq{func-der}. After some algebra it reproduces
Eq. \eq{eom-gauge} where the following identities were necessary
\begin{eqnarray}\label{zero1}
    && \frac{1}{2} \theta^{\alpha\beta} (\partial_\mu
    F_{\alpha\beta}) F^{\mu\nu}
    + \theta^{\mu \alpha} (\partial_\mu F_{\alpha\beta})
    F^{\beta \nu}= 0, \\
    \label{zero2}
    && -  \frac{1}{4}\theta^{\mu\nu} \partial_\mu
    (F_{\alpha\beta} F^{\alpha\beta}) + \theta^{\nu \alpha} (\partial_\mu
    F_{\alpha\beta}) F^{\beta \mu} = 0.
\end{eqnarray}
This proves the validity of the identity \eq{add-sw-eom} at least
to $\CO(\theta)$.

The above analysis shows that consistent results are obtained
irrespectively of whether the SW map is directly applied to the NC
action or on the NC object obtained from the NC action. For the
equation of motion, however, there is some subtlety which is next
discussed.

An application of the SW map on the equation of motion
\eq{lee-eom} yields, on using Eqs. \eq{eom-gauge} and \eq{current}, the
result
\begin{eqnarray} \label{tech-eom}
\widehat{W}^\nu & \equiv &
\partial_\mu \Bigl[ (1- \frac{1}{2}\theta^{\alpha\beta} F_{\alpha\beta}) F^{\mu \nu}
- (\theta F^2)^{\mu \nu} - (F \theta F)^{\mu \nu} \Bigr]
+ \theta^{\alpha\beta} \partial_\alpha \Bigl(A_\beta (\partial_\mu
F^{\mu\nu} + J^\nu) \Bigr) \xx
&& + J^\nu + (\theta F J)^\nu = 0.
\end{eqnarray}
On the other hand, the equation of motion \eq{lee-gi-eom} obtained
after applying the map on the NC action \eq{lee-ncqed} is given by
\begin{equation}\label{tech-deom}
W^\nu \equiv \partial_\mu \Bigl[ (1- \frac{1}{2}\theta^{\alpha\beta} F_{\alpha\beta}) F^{\mu \nu}
 - \frac{1}{4}\theta^{\mu\nu}F_{\alpha\beta}F^{\alpha\beta}
 - (\theta F^2)^{\mu \nu} - (F \theta F)^{\mu \nu} - (F^2 \theta)^{\mu \nu} \Bigr]
 + J^\nu = 0.
\end{equation}
The two equations \eq{tech-eom} and
\eq{tech-deom} are not identical, leading to the suspicion that
the implementation of the map is not a commutative operation
\ct{kruglov,grimstrup}. Why this difference occurs is not
difficult to understand. The equation \eq{tech-eom} was obtained
from a gauge covariant equation of motion \eq{lee-eom} while Eq.
\eq{tech-deom} was obtained from a gauge invariant one in
Eq. \eq{lee-gi-eom}. Nevertheless it is possible to establish a
compatibility by calculating the difference
\begin{equation}\label{diff}
    \widehat{W}^\nu - W^\nu = \theta^{\alpha\beta} \partial_\alpha
    \Bigl(A_\beta (\partial_\mu F^{\mu\nu} + J^\nu) \Bigr)
    + \theta^{\nu \alpha} F_{\alpha\beta} (\partial_\mu F^{\mu\beta} + J^\beta)
\end{equation}
which follows easily on using the identity (\ref{zero2}).  Now
it is seen from either Eq. \eq{tech-eom} or
Eq. \eq{tech-deom} that the term in the parenthesis $(\partial_\mu F^{\mu\nu} +
J^\nu)$ is at least of $\CO(\theta)$. Hence $\widehat{W}^\nu =
W^\nu$ up to the order we are dealing. This shows that the two
equations of motion are compatible.

Finally we would like to mention that ambiguities
\ct{asakawa,fidanza} in the basic SW map \eq{sw-a} do not affect the map
\eq{lee-gsw} among the currents. Any two solutions may differ
by a field dependent pure gauge $\p_\mu \Lambda(A)$ which is also
expected on general grounds since the SW transformation maps gauge
equivalent classes. Under this difference we find from
Eq. \eq{lee-gsw},
\begin{equation}
    \Delta \widehat{J}^\mu (x) = \int d^4 y J^\nu (y) \frac{\delta}{\delta \widehat{A}_\mu (x)}
    \Bigl(\p_\nu \Lambda(A)\Bigr) = -  \int d^4 y
    \p_\nu J^\nu (y) \frac{\delta \Lambda(A)}{\delta \widehat{A}_\mu
    (x)}= 0 \nonumber
\end{equation}
on using current conservation. Hence the map remains unchanged.
This is similar to the map \eq{sw-f} which is also unaffected \ct{fidanza}.

\subsection{Anomalies and the map}

The map for the currents found here also yields consistent results
even if the current is anomalous - that is, its usefulness is not
restricted to the strictly conserved or covariantly conserved
currents. We show this for leading order
in $\theta$. First note that Eq. \eq{current} can also be used to
relate the axial currents $\widehat{J}_5^\mu$ and $J_5^\mu$ at the
classical (tree) level. This is because, in that case, these
currents satisfy the same gauge transformation properties and
conservation laws as for the corresponding vector currents. The
issue is more subtle at the quantum level where, due to the one
loop effects, simultaneous conservation of $J^\mu$ and $J^\mu_5$
is not possible \ct{abj}. To fix our notions we take $J^\mu_5$ to
be anomalous. Since $\p_\mu J^\mu_5$ no longer vanishes, it is
natural to think that Eq. \eq{current} may be modified such
that it contains an extra $\CO(\theta)$-term, proportional to $\p_\mu
J^\mu_5$, in its right hand side. But, as long as we insist that
$\widehat{J}^\mu_5$ be $\star$-gauge covariant and $J^\mu_5$ be
gauge-invariant, the extra term should be gauge invariant by
itself. (In this regard, see Eq. \eq{gener-ans-j}). However, using
$\theta^{\alpha\beta}, \; F^{\mu\nu}$, and $\p_\nu J^\nu_5$, no
such gauge invariant term (with correct dimension and appropriate
tensor structure) can be found. Hence we expect our formula
\eq{current} to apply even for this anomalous case.

Given the relation \eq{current}, taking its covariant divergence
yields
\begin{equation}\label{a-claw}
 \widehat{D}_\mu \star \widehat{J}^\mu  = \partial_\mu
 J^\mu + \theta^{\alpha\beta} \partial_\alpha(A_\beta
 \partial_\mu J^\mu ).
\end{equation}
What was discussed till now $(\p_\mu J^\mu = \widehat{D}_\mu \star
\widehat{J}^\mu =0)$ is obviously compatible with the above relation.
Let us now consider the anomalous case (where, for notational
simplicity,
$\widehat{J}^\mu$ stands for axial current) for which we have \ct{anomaly}
\begin{equation}\label{nc-anomaly}
\widehat{D}_\mu \star \widehat{J}^\mu  = N \star (\widehat{F} \wedge
\widehat{F} \wedge \cdots \wedge \widehat{F})_{\rm{n-fold}}.
\end{equation}
The right hand side here is the (star) gauge covariant anomaly in
$d=2n$ dimensions, with $N$ being the normalization and using the
(star) wedge notation
\begin{equation}
    \star (\widehat{F} \wedge \cdots \wedge \widehat{F}) =
    \varepsilon_{\mu\nu \cdots \lambda\rho} \widehat{F}^{\mu\nu}
    \star \cdots \star \widehat{F}^{\lambda\rho}. \nonumber
\end{equation}
Up to $\CO(\theta)$ the star products involving $\widehat{F}$ can
be replaced by ordinary products so that, after applying the SW
map \eq{sw-f}, the anomaly \eq{nc-anomaly} reduces to,
\begin{equation} \label{sw-nc-anomaly}
\widehat{D}_\mu \star \widehat{J}^\mu \stackrel{\rm{SW \; map}}{=}
N \Bigl[(F \wedge F \wedge \cdots \wedge F) - n (F\theta F)
\wedge (F \wedge \cdots \wedge F)
- \theta^{\alpha\beta}A_\alpha \p_\beta(F \wedge
\cdots \wedge F) \Bigr].
\ee
Using the identity \ct{rabin},
\begin{equation}
\theta^{\alpha\beta}F_{\alpha\beta}(F \wedge
\cdots \wedge F) = -2n (F\theta F)
\wedge (F \wedge \cdots \wedge F), \nonumber
\end{equation}
we then get
\begin{equation}\label{sw-map-anomaly}
\widehat{D}_\mu \star \widehat{J}^\mu \stackrel{\rm{SW \; map}}{=}
N \Bigl[F \wedge F \wedge \cdots \wedge F
+ \theta^{\alpha\beta} \p_\alpha(A_\beta F \wedge
\cdots \wedge F) \Bigr].
\end{equation}
Comparing this with Eq. \eq{a-claw} the usual gauge invariant
anomaly in the SW deformed theory is deduced, i.e.,
\begin{equation}\label{d-anomaly}
\partial_\mu J^\mu = N (F \wedge F \wedge \cdots \wedge F)
\end{equation}
which is the expected result. Indeed the fact that the standard
ABJ-anomaly is not modified in $\theta$-expanded gauge theory was
earlier shown in \ct{brandt}. (For a mapping of the gauge
invariant anomaly in either description, see \ct{banerjee,rabin}.)
It appears, therefore, that our map \eq{current} correctly
incorporates quantum effects.

As another application, it is possible to discuss the shift in the
gauge invariance, exactly as happens in the commutative case.
Although it is possible, as before, to analyze in arbitrary
dimensions, we confine to $d=4$ where the usual ABJ-anomaly is
\begin{equation}\label{abj}
    \partial_\mu J^\mu = \frac{1}{16 \pi^2} \varepsilon_{\mu\nu \lambda\rho}
    F^{\mu\nu}F^{\lambda\rho}.
\end{equation}
Defining a modified current as
\begin{equation}\label{mod-current}
    \widetilde{J}^\mu = J^\mu - \frac{1}{8 \pi^2} \varepsilon^{\mu\nu \lambda\rho}
    A_\nu F_{\lambda\rho}
\end{equation}
leads to an anomaly free $(\partial_\mu \widetilde{J}^\mu=0)$ but
gauge noninvariant current \ct{bardeen}. To do a similar thing for the NC
case, rewrite the map \eq{current} by replacing $J^\mu$ in favor of
$\widetilde{J}^\mu$. The $\widetilde{J}^\mu$ independent terms are
then moved to the other side and a new
$\widehat{\widetilde{J}}^\mu$ is defined as
\begin{equation}\label{new-j}
    \widehat{\widetilde{J}}^\mu = \widehat{J}^\mu + \widehat{X}^\mu
    (\widehat{A}),
\end{equation}
so that
\begin{equation}\label{sw-newj}
 \widehat{\widetilde{J}}^\mu  =
 \widetilde{J}^\mu + (\theta F \widetilde{J})^\mu
 + \p_\alpha (\theta^{\alpha\beta} A_\beta \widetilde{J}^\mu).
\end{equation}
Note that all $A_\mu$-dependent terms lumped in $\widehat{X}^\mu$
can be recast in terms of $\widehat{A}_\mu$ using the SW map. Since
Eq. \eq{sw-newj} is structurally identical to Eq. \eq{current}, a
relation akin to Eq. \eq{a-claw} follows,
\begin{equation}\label{newj-claw}
 \widehat{D}_\mu \star \widehat{\widetilde{J}}^\mu  = \partial_\mu
 \widetilde{J}^\mu + \theta^{\alpha\beta} \partial_\alpha(A_\beta
 \partial_\mu \widetilde{J}^\mu ).
\end{equation}
Since $\partial_\mu \widetilde{J}^\mu = 0$ it follows that
$\widehat{D}_\mu \star \widehat{\widetilde{J}}^\mu  = 0$.
We are thereby successful in constructing an anomaly free current
which however does not transform covariantly. Its lack of
covariance is caused by the $\widehat{X}^\mu$ term in Eq.
\eq{new-j}, which plays a role analogous to the Chern-Simons three
form in the usual commutative description.

\section{Energy-Momentum Tensors and Lorentz Force Law}

The problems of defining EM tensors in NC gauge theories have been
studied by various authors
\ct{dorn,kruglov,grimstrup,das,ghosh} but the results have not always agreed.
In this section a systematic presentation is done which naturally
leads to a map among these tensors in the different (NC and
commutative) descriptions. A fall out of the analysis is the
Lorentz force law in NC space. As usual, the Lorentz force is
identified through considering the 4-divergence of electromagnetic
EM tensor.

To define a manifestly symmetric electromagnetic EM tensor on NC
space, the NC gauge fields are formally coupled to a weak external
gravitational field
\begin{eqnarray} \label{grav-ncqed}
    \widehat{S}_{\widehat{g}} &=&  -\frac{1}{4}
    \int d^4 x \sqrt{-\widehat{g}} \star \widehat{g}^{\mu \lambda}
    \star \widehat{g}^{\nu \rho}
    \star \widehat{F}_{\mu\nu} \star \widehat{F}_{\lambda\rho}.
\end{eqnarray}
The EM tensor is defined as
\begin{equation}\label{energy-momentum}
    \widehat{T}_{\mu\nu} = \frac{2}{\sqrt{-\widehat{g}}}
    \frac{\delta \widehat{S}_{\widehat{g}}}
    {\delta \widehat{g}^{\mu\nu}}|_{\widehat{g}^{\mu\nu} =
    \eta^{\mu\nu}}.
\end{equation}
There may be an ordering ambiguity in the above manipulation, but
that is inconsequential since eventually the metric is set flat.
We find
\begin{equation}\label{emt}
\widehat{T}_{\mu\nu} =
\frac{1}{2}(\widehat{F}_{\mu\lambda} \star \widehat{F}^\lambda_{\;\;\, \nu}
+ \widehat{F}_{\nu\lambda} \star \widehat{F}^\lambda_{\;\;\,\mu}) + \frac{1}{4}
\eta_{\mu\nu} \widehat{F}_{\lambda \rho} \star \widehat{F}^{\lambda
\rho}.
\end{equation}
This tensor is both symmetric and traceless. However it is not
star gauge invariant. Rather, it is star gauge covariant.
Expectedly, a covariant conservation law holds,
\begin{equation}\label{ent-conserve}
    \widehat{D}_\mu \star \widehat{T}^{\mu\nu} = 0
\end{equation}
which follows on using the source free equation of motion (see
Eq. \eq{lee-eom}) and the (NC) Bianchi identity
\begin{equation}\label{bianchi}
    \widehat{D}_\mu \star \widehat{F}_{\nu \lambda}
    +  \widehat{D}_\nu \star \widehat{F}_{\lambda \mu}
    +  \widehat{D}_\lambda \star \widehat{F}_{\mu \nu} = 0.
\end{equation}

Now the EM tensor $T_{\mu\nu}$ in commutative space is gauge
invariant and satisfies the ordinary conservation law. From an
algebraic point of view, therefore, $\widehat{T}_{\mu\nu}$ and
$T_{\mu\nu}$ (for each given $\nu$) simulate exactly the roles of the sources
$\widehat{J}^\mu$ and $J^\mu$. It is not unreasonable to expect
that the EM tensors also satisfy a map analogous to Eq.
\eq{current}, i.e., up to $\CO(\theta)$,
\begin{equation}\label{sw-em}
    \widehat{T}^{\mu\nu} = T^{\mu\nu} + (\theta F T)^{\mu\nu} +
    \p_\alpha (\theta^{\alpha\beta}A_\beta T^{\mu\nu}).
\end{equation}
We now prove that this is indeed so, simultaneously fixing the
structure of $T^{\mu\nu}$.

Before proceeding further it may be pointed out that Eq. \eq{emt}
also follows from a Noether procedure involving the combination of
translations with field dependent gauge transformations
\ct{jackiw}. Explicitly, acting the generator
\begin{equation}\label{improved}
    \widehat{W}_\mu^T = \frac{1}{2} \int d^4 x (\widehat{F}_{\mu \nu} \star
    \frac{\delta}{\delta\widehat{A}_\nu}
    +  \frac{\delta}{\delta\widehat{A}_\nu} \star \widehat{F}_{\mu \nu})
\end{equation}
on the flat NC action $\widehat{S}_{\rm{flat}}$ gives rise to
\begin{eqnarray}\label{ward}
\widehat{W}_\mu^T \widehat{S}_{\rm{flat}} &=& - \int d^4 x \widehat{D}^{\nu} \star
    \widehat{T}_{\mu\nu} \nonumber \\
    &=& - \int d^4 x \widehat{D}^{\nu} \star \Bigl(
\frac{1}{2}(\widehat{F}_{\mu\lambda} \star \widehat{F}^\lambda_{\;\;\, \nu}
+ \widehat{F}_{\nu\lambda} \star \widehat{F}^\lambda_{\;\;\,\mu}) + \frac{1}{4}
\eta_{\mu\nu} \widehat{F}_{\lambda \rho} \star \widehat{F}^{\lambda
\rho} \Bigr),
\end{eqnarray}
where we have used the identity (\ref{bianchi}).

Now expanding the EM tensor in Eq. (\ref{emt}) up to the leading order in
$\theta$, by using Eq. \eq{sw-f}, yields
\begin{eqnarray}\label{sw-emt}
\widehat{T}_{\mu\nu}|_{\rm{SW \; map}} &=&
 (1- \frac{1}{2} \theta^{\alpha\beta}F_{\alpha\beta})
 \Bigl( (F^2)_{\mu\nu} +  \frac{1}{4}\eta_{\mu\nu} F^2 \Bigr)
- (F^2 \theta F + F \theta F^2)_{\mu \nu}
+ \frac{1}{2} \eta_{\mu\nu} {\rm Tr} (F \theta F^2) \nonumber \\
&& + \partial_\alpha \Bigl[\theta^{\alpha\beta} A_\beta
\Bigl((F^2)_{\mu\nu} +  \frac{1}{4}\eta_{\mu\nu} F^2 \Bigr)
\Bigr] \nonumber \\
&=&\Bigl[(1- \frac{1}{2} \theta^{\alpha\beta}F_{\alpha\beta})
 F_{\mu\lambda} -  \frac{1}{4}\theta_{\mu \lambda} F^2
- (F^2 \theta + F \theta F + \theta F^2)_{\mu \lambda} \Bigr]
F^\lambda_{\;\;\, \nu} - \eta_{\mu\nu} \CL_{{\rm ph}}
\nonumber \\
&& +  (\theta F^3)_{\mu\nu} + \frac{1}{4} (\theta F)_{\mu\nu} F^2
+ \partial_\alpha (\theta^{\alpha\beta} A_\beta T^0_{\mu\nu}) \nonumber \\
&=& \Pi_{\mu\lambda} F^\lambda_{\;\;\, \nu}
- \eta_{\mu\nu} \CL_{{\rm ph}} + ( \theta F T^0)_{\mu\nu}
+ \partial_\alpha (\theta^{\alpha\beta} A_\beta T^0_{\mu\nu})
\end{eqnarray}
where $\CL_{{\rm ph}}$ is the Lagrangian density for nonlinear
photons read off from Eq. \eq{lee-dqed} and $ T^0_{\mu\nu}$ is the
EM tensor for $\theta = 0$,
\begin{equation}\label{emt-0}
    T^0_{\mu\nu} = (F^2)_{\mu\nu} +  \frac{1}{4}\eta_{\mu\nu} F^2
\end{equation}
while $ \Pi_{\mu\nu}$ is the generalized canonical momenta as defined by
\begin{equation}\label{can-mon}
    \Pi_{\mu\nu} =  - \frac{\partial \mathcal{L}_{{\rm ph}} }{\partial (\partial^\mu
    A^\nu)} = (1- \frac{1}{2} \theta^{\alpha\beta}F_{\alpha\beta})
 F_{\mu\nu} -  \frac{1}{4}\theta_{\mu \nu} F^2
- (F^2 \theta + F \theta F + \theta F^2)_{\mu \nu}.
\end{equation}

The EM tensor in the commutative picture is likewise obtained from
the operator analogous to that in Eq. \eq{improved}, i.e., using
the generator \ct{jackiw}
\be \la{improved-c}
W_\mu^T = \int d^4 x F_{\mu\nu} \frac{\delta}{\delta A_\nu},
\ee
and the relation
\be \la{jackiw-c}
W_\mu^T S_{{\rm ph}} = -  \int d^4 x \p^\nu T_{\mu\nu}
\ee
where $S_{{\rm ph}}$ is defined in Eq. \eq{lee-dqed}, so that we
find
\begin{equation}\label{inv-emt}
    T_{\mu\nu} = \Pi_{\mu\lambda} F^\lambda_{\;\;\, \nu}
- \eta_{\mu\nu} \CL_{{\rm ph}}.
\end{equation}
Using the free field equation of motion $\p^\mu \Pi_{\mu\nu} = 0$
(which follows from Eq. \eq{tech-deom} by setting $J_\nu = 0$) and
\begin{equation}\label{identity-charge}
    \partial^\nu \CL_{{\rm ph}} -
    \frac{\partial \CL_{{\rm ph}} }{\partial (\partial_\mu A_\lambda)}
 \partial^\nu ( \partial_\mu A_\lambda) = 0,
\end{equation}
it is easy to see that
\begin{equation}\label{free-claw}
\partial^\mu T_{\mu\nu} = 0.
\end{equation}
Since this EM tensor was obtained from the commutative equivalent
of the NC theory, it is the one that should be used in the map.
Furthermore it is reassuring to note that $T_{\mu\nu}$ is both
gauge invariant and conserved, exactly as desired. Now
$T^0_{\mu\nu}$ in Eq. \eq{emt-0} and $T_{\mu\nu}$ in Eq. \eq{inv-emt} differ by
terms of $\CO(\theta)$, so that Eq. \eq{sw-emt} may be cast
precisely in the form \eq{sw-em}. This completes the derivation, up to
$\CO(\theta)$, of the map between the EM tensors.
Note that $T^{\mu\nu}$ appearing in the map
is neither symmetric nor traceless. This is due to the fact that
Lorentz and classical conformal invariance are broken in NC
electrodynamics \ct{carroll}.

Inclusion of sources does not pose any problem. The structures of
the relevant electromagnetic EM tensors remain the same, but the
conservation laws in Eqs. \eq{ent-conserve} and \eq{free-claw}
are modified leading to the respective Lorentz force laws.

Starting from Eq. \eq{emt} and using the equation of motion \eq{lee-eom}
together with the Bianchi identity \eq{bianchi} immediately yields
the NC generalization of the Lorentz force law
\begin{equation}\label{nc-force}
\widehat{D}_\mu \star \widehat{T}^{\mu\nu} = - \frac{1}{2} (
\widehat{J}_\mu \star \widehat{F}^{\mu\nu} + \widehat{F}^{\mu\nu}
\star \widehat{J}_\mu ).
\end{equation}
Similarly, the corresponding law in the commutative picture
emerges by considering the equation of motion \eq{lee-gi-eom} and
takes the form
\begin{equation}\label{dqed-claw}
\partial_\mu T^{\mu\nu} = - J_\mu F^{\mu\nu}.
\end{equation}
As a consistency of the whole program, we show that the
deformation in the Lorentz force law  as given by Eq. \eq{nc-force}
is compatible with Eq. \eq{dqed-claw}. Using the expressions for the various maps,
it turns out that, up to $\CO(\theta)$,
\begin{equation} \label{sw-nceom}
\widehat{D}_\mu \star \widehat{T}^{\mu\nu}|_{\rm{SW \; map}}
= \partial_\mu T^{\mu\nu}
+ \partial_\alpha (\theta^{\alpha\beta} A_\beta \partial_\mu T^{\mu\nu})
\end{equation}
and,
\begin{equation}\label{sw-ncforce}
 \frac{1}{2} (\widehat{J}_\mu \star \widehat{F}^{\mu\nu} + \widehat{F}^{\mu\nu}
\star \widehat{J}_\mu )|_{\rm{SW \; map}} = J_\mu
F^{\mu\nu} + \partial_\alpha (\theta^{\alpha\beta} A_\beta J_\mu
F^{\mu\nu} ).
\end{equation}
Adding them together yields,
\begin{eqnarray}\label{sw-nclorentz}
  && \Bigl[ \widehat{D}_\mu \star \widehat{T}^{\mu\nu} +
 \frac{1}{2} (\widehat{J}_\mu \star \widehat{F}^{\mu\nu} + \widehat{F}^{\mu\nu}
\star \widehat{J}_\mu )\Bigr]_{\rm{SW \; map}} = \xx
&& \hspace{4cm} \partial_\mu T^{\mu\nu} + J_\mu F^{\mu\nu}
+ \partial_\alpha \Bigl(\theta^{\alpha\beta} A_\beta
(\partial_\mu T^{\mu\nu} + J_\mu F^{\mu\nu})\Bigr).
\end{eqnarray}
It is now clear that Eq. \eq{dqed-claw} implies Eq. \eq{nc-force}.
Incidentally Eq. \eq{sw-nceom} is the exact analogue of Eq.
\eq{a-claw} that maps the source divergence.

\section{Discussion}

We have provided a Seiberg-Witten like map relating the sources in
the noncommutative (NC) and commutative descriptions. With its
help, a commutative equivalent of NC electrodynamics with sources
was formulated. Consistent results were obtained by applying the
map either on the action or on the equations of motion. Although
the map could, in principle, be worked to higher orders
in $\theta$ (the NC parameter), for reasons of compactness $\CO(\theta)$
results were explicitly analyzed. In this regime the map was also
used to relate the star gauge covariant anomaly in the NC theory
with the gauge invariant ABJ-anomaly in the $\theta$-deformed
theory.

Our methods were then extended to reveal a mapping among the
energy-momentum (EM) tensors in the two descriptions. In the
presence of sources, the NC generalization of the Lorentz force
law was derived. The various maps were used to show that the
deformation of Lorentz force law was consistent in the sense that
enforcing this law in the commutative picture automatically
enforced it in the NC picture.

Despite the different methods and different variables (e.g.
currents, EM tensors, etc) used, a universal structure seemed to
emerge in the various maps, at least to $\CO(\theta)$. This
reinforces the role of gauge transformations in mapping variables
in NC gauge theories with their commutative equivalents.

As yet another manifestation of this universality, we discuss, for
the special case of two dimensions, a Sugawara-type construction
where EM tensors are expressed in terms of currents.
In two dimensions the NC parameter $\theta^{\mu\nu} = \theta
\varepsilon^{\mu\nu}$ really transforms as a Lorentz tensor so
that invariances or symmetries not valid in higher dimensions may
be restored in this case. This leads to a viability of alternative
formulations where the EM tensor is symmetric. It may be recalled
that even in commutative field theory, two dimensions play a
special role with properties like exact solvability, bosonization,
etc.

We begin with the commutative theory. Here it is known
\ct{abdalla} that the EM tensor of a conformally invariant theory
is expressed solely in terms of the currents,
\begin{equation}\label{suga-emt-c}
    T_{\mu\nu}= \frac{\pi}{2}(J_\mu J_\nu + J_\nu J_\mu -
    \eta_{\mu\nu} J_\lambda J^\lambda)
\end{equation}
which is referred as the Sugawara form. Then, in the NC theory
context, we may consider a natural noncommuatative generalization
of this form, i.e.,
\begin{equation}\label{suga-emt-nc}
    \widehat{T}_{\mu\nu} = \frac{\pi}{2}(\widehat{J}_\mu \star
    \widehat{J}_\nu + \widehat{J}_\nu  \star \widehat{J}_\mu -
    \eta_{\mu\nu} \widehat{J}_\lambda \star \widehat{J}^\lambda).
\end{equation}
Now the EM tensor of the commutative equivalent of this NC theory
can be obtained using our map \eq{sw-em}, together with the
current map \eq{current}. A surprise is that, for this EM tensor,
we find back the form \eq{suga-emt-c}; but, of course,
$J^\mu$ can contain $\theta$-dependent corrections here.
This is demonstrated below.

Expanding the star product in Eq. \eq{suga-emt-nc} yields,
\begin{equation}\label{suga-sw-emt1}
    \widehat{T}_{\mu\nu} = \frac{\pi}{2}(\widehat{J}_\mu
    \widehat{J}_\nu + \widehat{J}_\nu  \widehat{J}_\mu -
    \eta_{\mu\nu} \widehat{J}_\lambda \widehat{J}^\lambda)
    + \frac{i \pi}{4} \theta^{\alpha\beta}(\p_\alpha \widehat{J}_\mu
    \p_\beta \widehat{J}_\nu + \p_\alpha \widehat{J}_\nu
    \p_\beta \widehat{J}_\mu).
\end{equation}
In the second parentheses, the NC variable can be replaced by the
commutative one, since the analysis is done up to $\CO(\theta)$.
Then it can be expressed as a commutator $\theta^{\alpha\beta}
\p_\alpha \p_\beta [J_\mu , J_\nu]$\footnote{Actually all
products of currents have to be properly interpreted by a
point-splitting regularization \ct{abdalla} in which case
$[J_\mu(x), J_\nu(y)]$ is just a function of $(x-y)$. Indeed, to
give a definite meaning to the Sugawara construction, such a
prescription is implicitly assumed.} which vanishes from symmetry
arguments. Now inserting the map \eq{current} in Eq.
\eq{suga-sw-emt1} leads to,
\begin{equation}\label{suga-sw-emt2}
\widehat{T}_{\mu\nu} = T_{\mu\nu} + \rm{order} \; \theta \; \rm{terms}
\end{equation}
where $T_{\mu\nu}$ is defined in Eq. \eq{suga-emt-c}. After a
slightly lengthy algebra, we get
\begin{equation}\label{suga-sw-emt3}
\widehat{T}_{\mu\nu} = T_{\mu\nu} + 2(\theta F T)_{\mu\nu}
+ \theta^{\alpha\beta} F_{\alpha\beta}T_{\mu\nu} + \theta^{\alpha\beta}
A_\beta \p_\alpha T_{\mu\nu}.
\end{equation}
Using the identity,
\begin{equation}\label{suga-identity}
(\theta F T)_{\mu\nu} = \theta_{\mu\alpha} F^{\alpha\beta}
T_{\beta\nu} = - \half \theta^{\alpha\beta} F_{\alpha\beta}
T_{\mu\nu},
\end{equation}
the equation \eq{suga-sw-emt3} then reduces to
\begin{equation}\label{suga-sw-final}
\widehat{T}_{\mu\nu} = T_{\mu\nu} + (\theta F T)_{\mu\nu}
 + \p_\alpha (\theta^{\alpha\beta} A_\beta T_{\mu\nu}).
\end{equation}
Since this has an identical structure as Eq. \eq{sw-em},
we now conclude that $T^{\mu\nu}$ as given by Eq. \eq{suga-emt-c}
is the full expression to $\CO(\theta)$.
Incidentally, contrary to the
earlier case, here both $\widehat{T}_{\mu\nu}$ and $T_{\mu\nu}$
are symmetric because $\theta^{\mu\nu}$ in two dimensions
is invariant under Lorentz transformations.
Also it appears that, at least to order $\theta$, the scale invariance is preserved.

Finally, from a general point of  view, we end with the following
remarks: the fact that anomalies could be related (Section 2.3),
strongly suggests the feasibility of obtaining SW-type maps for
effective actions. These would find an obvious application of
connecting consistent as well as covariant anomalies for $U(N)$
gauge theories in the two descriptions. Presumably trace anomalies
related to the EM tensors could also be discussed within this
formulation. These topics are left for the future.

\section*{Acknowledgments}

The work of RB was supported by funds from Sungkyunkwan University
and Seoul National University. He is also grateful to the members
of the respective physics departments for their gracious
hospitality. The work of CL was supported by Korea Science
Foundation ABRL program (R14-2003-012-01002-0) and by a 2003
Interdisciplinary Research Grant of Seoul National University. HSY
was supported by the Brain Korea 21 Project in 2003.


\bibliographystyle{JHEP-like}

\end{document}